\documentclass[lettersize,journal]{IEEEtran}
\usepackage{amsmath,amsfonts}
\usepackage{algorithmicx} \usepackage{algpseudocode}
\usepackage{algorithm}
\usepackage{array}
\usepackage[caption=false,font=normalsize,labelfont=sf,textfont=sf]{subfig}
\usepackage{textcomp}
\usepackage{stfloats}
\usepackage{url}
\usepackage{verbatim}
\usepackage{graphicx}
\usepackage{cite}
\usepackage{multirow} 
\usepackage{booktabs} 
\usepackage{makecell}
\usepackage{algpseudocode}
\usepackage{xcolor}  
\usepackage{stfloats}
\usepackage{lettrine}
\usepackage{orcidlink}

\begin{document}

\title{Generative Model-Assisted Demosaicing for Cross-multispectral Cameras}
\author{
Jiahui Luo$^*{}^{\orcidlink{0009-0006-7221-0068}}$, 
Kai Feng$^*{}^{\orcidlink{0000-0002-3066-0056}}$, 
Haijin Zeng$^\dagger{}^{\orcidlink{0000-0003-0398-3316}}$, 
Yongyong Chen$^\dagger{}^{\orcidlink{0000-0003-1970-1993}}$,~\IEEEmembership{Member,~IEEE}
\thanks{This work was supported by the National Natural Science Foundation of China under Grant 62106063. (Jiahui Luo and Kai Feng contributed equally to this work. Corresponding authors: Yongyong Chen; Haijin Zeng).}

\thanks{Jiahui Luo and Yongyong Chen are with the School of Computer Science and Technology, Harbin Institute of Technology (Shenzhen),
 Shenzhen 518055, China (e-mail: 23s151054@stu.hit.edu.cn; YongyongChen.cn@gmail.com).}
\thanks{Kai Feng is with the School of Automation, Northwestern Polytechnical University, Xi’an 710072, China (e-mail: 2018100620@mail.nwpu.edu.cn).}
\thanks{Haijin Zeng is with the Image Processing and Interpretation, IMEC Research Group, Ghent University, 9000 Ghent, Belgium (e-mail: Haijin.Zeng@UGent.be).}
}


\maketitle

\begin{abstract}
As a crucial part of the spectral filter array (SFA)-based multispectral imaging process, spectral demosaicing has exploded with the proliferation of deep learning techniques. However, (1) bothering by the difficulty of capturing corresponding labels for real data or simulating the practical spectral imaging process, end-to-end networks trained in a supervised manner using simulated data often perform poorly on real data. (2) cross-camera spectral discrepancies make it difficult to apply pre-trained models to new cameras. (3) existing demosaicing networks are prone to introducing visual artifacts on hard cases due to the interpolation of unknown values. To address these issues, we propose a hybrid supervised training method with the assistance of the self-supervised generative model, which performs well on real data across different spectral cameras. Specifically, our approach consists of three steps: (1) Pre-Training step: training the end-to-end neural network on a large amount of simulated data; (2) Pseudo-Pairing step: generating pseudo-labels of real target data using the self-supervised generative model; (3) Fine-Tuning step: fine-tuning the pre-trained model on the pseudo data pairs obtained in (2). To alleviate artifacts, we propose a frequency-domain hard patch selection method that identifies artifact-prone regions by analyzing spectral discrepancies using Fourier transform and filtering techniques, allowing targeted fine-tuning to enhance demosaicing performance. Finally, we propose UniSpecTest, a real-world multispectral mosaic image dataset for testing. Ablation experiments have demonstrated the effectiveness of each training step, and extensive experiments on both synthetic and real datasets show that our method achieves significant performance gains compared to state-of-the-art techniques. 
\end{abstract}


\begin{figure}[!ht]
\centering
\includegraphics[width=0.47\textwidth]{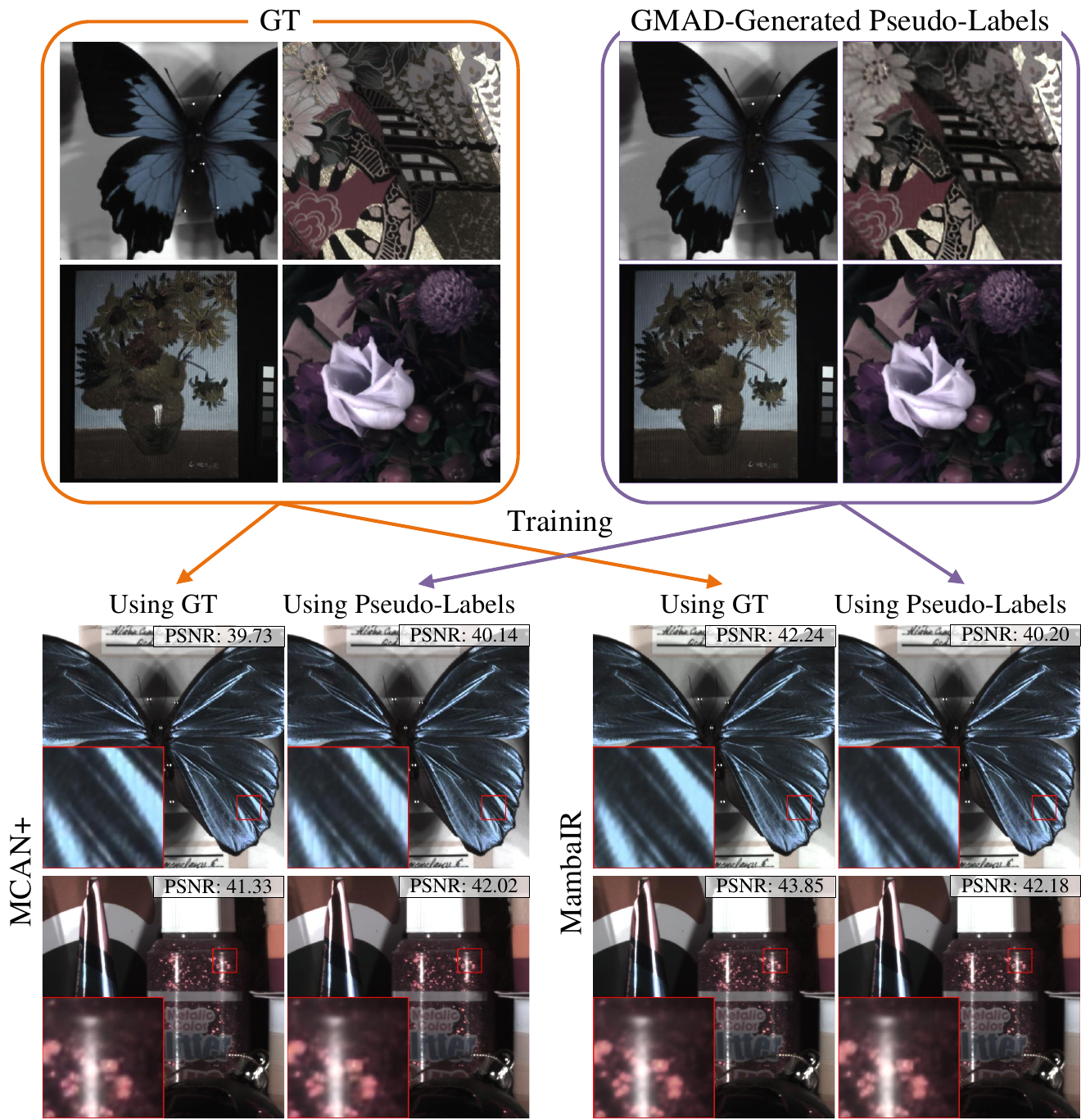} 
\caption{Comparison of training performance using GMAD-generated pseudo-labels versus GT labels across two backbone networks (MCAN+ and MambaIR). PSNR evaluation demonstrates that on the MCAN+ network, training with GMAD-generated pseudo-labels even outperforms GT-based training, thereby validating that the proposed GMAD method can achieve competitive performance when GT data is unavailable.}
\label{DIPorGT_teaser_figure}
\end{figure}

\section{Introduction}
\lettrine[lines=2]{C}{OMPARED} to RGB images, multispectral images (MSI) incorporate extra one-dimensional spectral information that captures image data at multiple spectral frequencies and can non-destructively identify the composition of substances, which makes them important for applications in areas such as remote sensing \cite{satellite}, environmental monitoring \cite{LuDLHS20}, medical diagnostics \cite{Medical} and food inspection \cite{food_application}. Methods for obtaining multispectral images include scanning spectral imaging and snapshot spectral imaging \cite{deep_review}. MSI can be obtained through scanning spectral imaging techniques, such as whiskbroom, pushbroom, and wavelength scanning. These techniques acquire two-dimensional data per cycle and thus require multiple measurements to synthesize a full three-dimensional data cube. This limitation makes it challenging to capture accurate data for dynamic scenes. To address these issues, snapshot spectral cameras equipped with spectral filter arrays (SFAs) \cite{SFA1, TT31} have been introduced to collect undersampled 3D data cubes in a single exposure. However, these cameras rely on multispectral demosaicing to reconstruct missing spectral information and generate the full data cube. Demosaicing is therefore crucial for ensuring accurate spectral recovery, as the quality of the process directly influences subsequent analyses and real-world applications \cite{2015snapshot, 2017snapshot} that rely on reliable MSI data.

Deep learning has become an essential high performance tool for demosaicing. The commonly utilized algorithms \cite{CNN1, CNN2, CNN3, PPID, SpNet, InNet} are based on convolutional neural network (CNN), which effectively incorporate additional constraints that are meaningful during the learning process, such as spatial invariance and resilience to minor rotations and deformations. More recently, researchers have adopted the transformer architecture for demosaicing \cite{Transformer1, Transformer2, Transformer3}, leveraging non-local similarities to enhance demosaicing performance. Compared to CNN architectures, transformers require a larger volume of training data. However, these methods often fall short in real-world applications due to the absence of ground truth (GT) \cite{lackGT}, leading to suboptimal image reconstruction accuracy.

A common strategy is to solve this problem by large scale pre-training \cite{windrim2018pretraining}. However, due to illumination, spectral sensitivity differences of different cameras, and non-uniform SFA standards across vendors \cite{9bands}, there are variations in the spectral features of images captured from the same object in different scenes, which leads to poor results in the pre-trained models when processing new camera data.
Other end-to-end neural networks, such as MCAN \cite{MCAN}, employ an adaptive feature extraction attention mechanism to enhance the demosaicing results. Despite their advantages, such supervised methods cannot be directly applied to new cameras lacking GT. To address the challenge of limited multispectral data, numerous unsupervised methods have been developed that rely on image priors. For instance, traditional methods like the WB \cite{WB} and ItSD \cite{ItSD} utilize bilinear and iterative interpolation to exploit spatial and inter-channel correlations, respectively. However, these techniques often result in blurred boundaries in areas with complex textures or edges. Advanced approaches such as DIP \cite{DIP} and USD \cite{USD} leverage image self-similarity and iterative strategies, respectively, to enhance image reconstruction. Despite their promise, these two methods are notably time-intensive and require careful parameter tuning. Moreover, inaccurate priors and insufficient utilization of correlations are prone to introducing artifacts, as shown in Fig. \ref{Curve_figure}(a). These artifacts mainly result from the interpolation of unknown values \cite{SpNet} and flaws in spectral filter design \cite{diaz2023SFAdesign}.

Generative model has already demonstrated its power in tasks such as image denoising \cite{generative_denoise}, image super-resolution \cite{generative_SR}, and image generation \cite{generative_restoration}. By learning the intrinsic distribution of data, these models can produce demosaiced labels that are highly close to real data. 
Building on this capability, we propose a \textbf{G}enerative \textbf{M}odel-\textbf{A}ssisted  \textbf{D}emosaicing (GMAD) to facilitate supervised training in real-world scenarios that lack GT, which includes a fine-tuning method to bridge the gap between different datasets. As shown in Fig. \ref{DIPorGT_teaser_figure}, even without using GT data, the results obtained using GMAD training are comparable to those from the same model trained with GT. In summary, GMAD comprises the following three primary steps:  
(1) \textbf{Pre-Training Step}. The end-to-end network is initially trained on a large simulated spectral dataset. This approach enhances the feature extraction capability of the network and reduces the risk of overfitting on real-world datasets.
(2) \textbf{Pseudo-Pairing Step}. In this step, we use the DIP to process mosaic images from the SFA camera to generate pseudo demosaiced cubes. The DIP is capable of generating images from a network of randomly initialized encoders and decoders. This solves the challenge of adapting the model to a new camera in the absence of GT.
(3) \textbf{Fine-Tuning Step}. The pre-trained model is then fine-tuned to adapt to the parameters of a different multispectral camera using a pseudo-paired dataset.

To mitigate artifacts in the demosaicing process, we propose an algorithm that filters out challenging patches in the frequency domain from the pseudo-paired dataset. This enables the model to focus on critical regions, such as image edges and complex textures, during step 3, thereby enhancing both the overall image quality and spectral accuracy.

In summary, the main contributions of this article are as follows:
\begin{itemize}
\item We propose GMAD, a spectral demosaicing training method. GMAD utilizes a generative model to accomplish demosaicing using a supervised network without labels. It not only improves the performance, but also extends the application of supervised learning in the field of demosaicing.
\item We developed a frequency-domain hard patch selection algorithm that significantly reduces demosaicing artifacts while enhancing spectral confidence.
\item To further validate our approach, we introduce the UniSpecTest dataset, a comprehensive and diverse benchmark designed for rigorous evaluation of spectral demosaicing performance.
\end{itemize}

The remainder of this paper is organized as follows. Section II reviews existing unsupervised and supervised spectral demosaicing methods. Section III presents a hybrid supervised training method and an algorithm to mitigate artifacts, as well as a new mosaic dataset. Section IV presents extensive performance evaluations and comparisons with state-of-the-art methods. Section V concludes the paper.

\section{Related Work}
Based on training method, multispectral demosaicing algorithms can be categorized into two main groups: unsupervised methods and supervised methods. Unsupervised methods utilize statistical information in multispectral images, such as sparsity, spatial structure similarity and spectral correlation. On the other hand, supervised methods are data-driven and well-suited for managing the complexity of high-dimensional data.

\subsection{Unsupervised Multispectral Demosaicing}
Conventional unsupervised demosaicing methods employ interpolation algorithms that utilize intra-channel correlation and inter-channel correlation to recover the missing band information for each pixel. 
For example, the WB \cite{WB} performs bilinear interpolation through spatial correlation in each channel, and the ItSD \cite{ItSD} enhances the inter-channel correlation through repeated interpolation. However, the interpolation technique cannot well deal with the local regions of the image with drastic jumps, such as edges, textures and other detail information, thus leading to a blur of the image edges. 
Thereafter, PPID \cite{PPID} obtains pseudo-panchromatic images by calculating the average of all spectral channels to refine the edges and texture parts. 
However, training the network solely with unsupervised loss functions imposes weak constraints, making it challenging to recover the substantial missing information in the demosaiced images.
DIP is a self-supervised generative network, which relies only on the structure of the network and the prior of inner images. 
USD employs a self-correcting iterative strategy, where its demosaiced cubes are resampled to create new spectral mosaic images for the next input. Thus, the cubes generated in previous iterations guide the demosaicing in subsequent cycles. Nonetheless, both approaches necessitate extended training durations and demand precise parameter tuning.

\subsection{Supervised Multispectral Demosaicing}
Supervised neural network methods depend on labeled training data to refine their accuracy. SpNet \cite{SpNet} applies a standard convolution directly to a mosaic cube of uninterpolated information to be recovered, but the sparse input is not conducive to neural network training. Thereafter, InNet inputs the multispectral data cube generated after bilinear interpolation into the network and proposes a three-dimensional convolutional layer to characterize the multispectral data cube. This approach improves the sharpness of the edges of the demosaiced image but produces more severe aliasing color artifacts. Further, MCAN proposes a convolution-attention mechanism for adaptive feature extraction, which improves the synthesis performance, but it also does not adequately address the generation of artifacts. It should be noted that while supervised deep neural networks work well, complex network models typically have more parameters and deeper structure, therefore requiring more data to learn general characteristics rather than noise \cite{zhang2022survey}. 

\section{The Proposed Method}
\subsection{Preliminary}

\begin{figure*}[ht]
\centering
\includegraphics[width=0.98\textwidth]{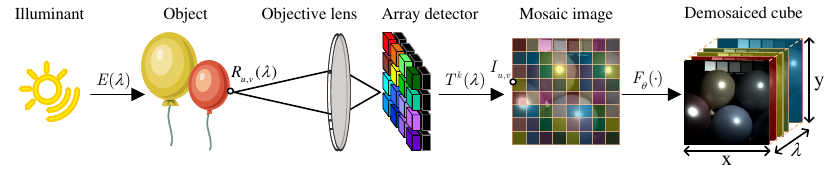} 
\caption{
The imaging process of a snapshot spectral imaging system. Numerous factors in real-world environments influence the generation of a mosaic image.
}
\label{Process}
\end{figure*}
The optical imaging process of a Snapshot Spectral Image (SSI) system based on an SFA can be explicitly described by the following equations \cite{huang2022spectral}.
First, the response of each spectral channel at a given pixel $(u,v)$ is obtained by integrating over the wavelength range:
\begin{equation}
    {I^k_{u,v}} = \int_{{\lambda _{\min }}}^{{\lambda _{\max }}} {E(\lambda ){R_{u,v}}(\lambda ){T^k}(\lambda )d\lambda },
\end{equation}
where $k={1,2,..., C}$ represents the number of imaging bands of the system. $E(\lambda )$ and ${R_{u,v}}(\lambda )$ denote the illumination spectrum and the spectral reflectance of the target object. ${T^k}(\lambda )$ is the transmission function of the $k$th filter in the SFAs.

This integral computes the cumulative contribution from all wavelengths within the interval $[\lambda _{\min }, \lambda _{\max }]$ for the $k$th channel.
The imaging process of this system is shown in Fig. \ref{Process}, and in a simplified form, the overall imaging process can be represented as:
\begin{equation} 
    A=Ha+n,
\end{equation}
where $A \in \mathbb{R}^{M\times N}$ is the 2D mosaic image directly acquired by the system sensor, and $a \in \mathbb{R}^{M \times N \times C}$ is the 3D spectral data cube of the target object, whose spatial resolution is determined by the resolution of the imaging system's sensor. The spectral resolution is determined by the mode of the spectral filter array, whose transmission matrix is $H$, and $n$ denotes the noise introduced by the system's imaging process.

The multispectral demosaicing algorithm is tasked with reconstructing $a$ from the captured $A$. In contrast to compressive spectral imaging, the inverse mosaic problem in SSI is more akin to the tensor complementation problem than the inverse convolution problem associated with compressive perception \cite{diaz2023SFAdesign}. In addition, in real-world scenarios, the functions of $E(\lambda )$ and ${R_{u,v}}(\lambda )$ are typically complex and variable, posing significant challenges for multispectral demosaicing.

\begin{figure*}[ht]
\centering
\includegraphics[width=1\textwidth]{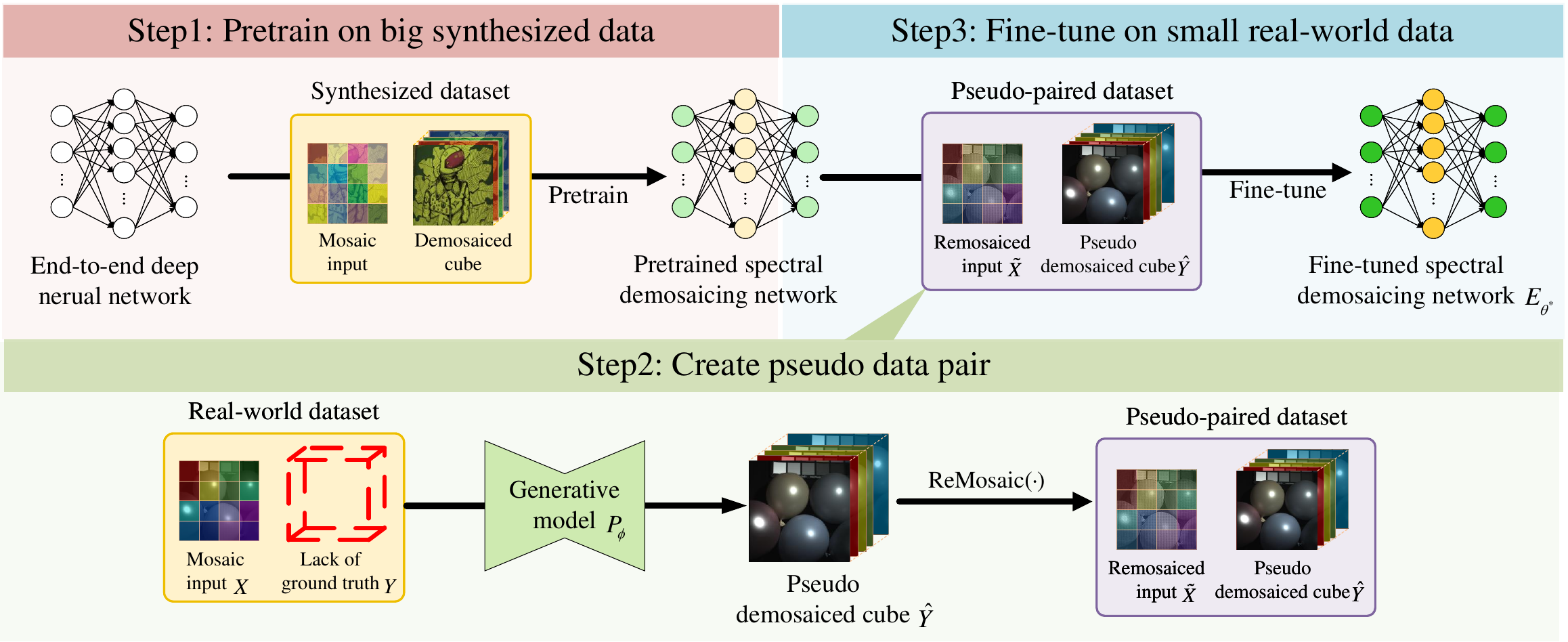} 
\caption{
Overview of our GMAD pipeline. \textbf{Step 1} involves pre-training the network on abundant simulated data to learn key multispectral image features. \textbf{Step 2} employs a generative model to generate a pseudo-demosaiced cube, generating a paired training dataset for the target domain that lacks GT. \textbf{Step 3} consists of a fine-tuning step, where supervised fine-tuning on the pseudo-paired dataset enhances network performance in real applications. This method combines unsupervised and supervised learning to enable knowledge transfer and model generalization across different SFA cameras without requiring GT.
}
\label{frame}
\end{figure*}

\subsection{Three-stage Training Method}
The overview of our proposed GMAD is shown in Fig. \ref{frame}. It can be divided into three steps, where the first and second steps can be performed in parallel. The two key points of this training method are: first, the pseudo labels of the target dataset are generated by a generative model, which enables the transition from unsupervised to supervised training and provides high quality demosaicing guidance; and second, the pre-training on a large number of simulated datasets reduces the difficulty of training and enhances the generalization of the model.

\subsubsection{Step 1: Pre-train on Large Simulated Data}
Many methods use pre-trained networks as initial weights to accelerate the learning process \cite{pretrain1, pretrain2, pretrain3}. Meanwhile, simulated data is much easier to obtain than real-world data. Thus, we adopt a strategy of pre-training the demosaicing network on a large simulated dataset.
Given a large simulated dataset ${\cal D}_{{\rm{sim}}}$, the first goal is to pre-train the model by minimizing the supervised loss function ${\cal L}_{{\rm{sup}}}$. Let ${F_\theta }:X \to Y$ denote the network function parameterized by $\theta$:
\begin{equation}
    {\theta _0} = \arg \mathop {\min }\limits_\theta  {\mathbb{E}_{(x,y) \in {{\cal D}_{{\rm{sim}}}}}}[{{\cal L}_{{\rm{sup}}}}({F_\theta }(x),y)],
\end{equation}
where $x$ denotes the input mosaic images and $y$ represents their corresponding labels.   
Pre-training on the large simulated dataset allows the network ${F_\theta }$ to learn general multispectral image features, providing a robust starting point for subsequent fine-tuning. Additionally, initializing the model with pre-trained parameters ${\theta _0}$ can significantly reduce the number of iterations required for convergence during fine-tuning \cite{pretrain1}.

\subsubsection{Step 2: Generate Pseudo-Paired Data}

For each real-world image $X \in \mathbb{R}^{H \times W}$ without GT, pseudo demosaiced cube $\hat Y \in \mathbb{R}^{H \times W \times C}$ is generated using a generative model $P_{\phi}$, where $\phi$ represents an image-specific parameters of the model:
\begin{equation}
    {\hat Y} = {P_{{\phi }}}(X).
\end{equation}
Here, $H$ and $W$ represent the height and width of the image, respectively. The number of spectral channels $C$ corresponds to $N_{MSFA}^2$, where $N_{MSFA}$ is the size of the SFA pattern. Note that varying $\phi$ (e.g., using different parameter sets $\phi_i$ ) can yield different pseudo demosaiced cubes for the same input $X$.

Next, a remosaicing operation ${\mathop{\rm Re}\nolimits} mosaic( \cdot )$ is applied to $\hat Y$ to form a pseudo-paired dataset ${\cal D}_{pseudo}$:
\begin{equation}
    {{\cal D}_{pseudo}} = \{ \tilde X,\hat Y\}  = \{ {\mathop{\rm Re}\nolimits} mosaic(\hat Y),\hat Y\}. 
\end{equation}
Consequently, the remosaiced image $\tilde X$ and the pseudo labels $\hat Y$ achieve a more precise alignment, which is preferable to directly using the original mosaic image $X$. Specifically, ${\mathop{\rm Re}\nolimits} mosaic( \cdot )$ is the process of simulating SFA camera sampling from 3D matrix data to obtain the corresponding 2D mosaic image using a spectral filter mask $M$. We can denote the row and column coordinates of a pixel as $u$ and $v$,  and define the mask $M$ as:
\begin{equation}
    M_{i,j}(u,v) = \mathbb{I} \big( u \bmod N_{\text{MSFA}} = i \land v \bmod N_{\text{MSFA}} = j \big),
\end{equation}
where $\mathbb{I}(condition)$ means “return 1 when the $condition$ is true, otherwise return 0”, $i,j \in \{ 0,1,...,N_{MSFA}^2 - 1\}$.
After that, for each band $k$ (corresponding to the position in the SFA $(i,j)$ of the filter), each pixel value of the mosaic image $\tilde X$ can be expressed as:
\begin{equation}
    \tilde X(u,v) = \sum\limits_{k = 1}^C {{{\hat Y}^k}(u,v) \cdot {M_{i,j}}(u,v)},
\end{equation}
where ${\hat Y^k}(u,v)$ denotes the pixel value of the pseudo demosaiced cube at band $k$ and pixel position $(u,v)$.

\subsubsection{Step 3: Fine-tune on Small Real-world Data}
The direct application of a model to a new target camera without fine-tuning often results in unsatisfactory performance, as shown in Fig. \ref{compare_Stereo} in the supervised method. For this reason, Step 3 uses the pseudo-paired data generated in Step 2 for fine-tuning.  
Specifically, supervised learning is performed on the pseudo-paired dataset ${\cal D}_{pseudo}$ to fine-tune the model parameters $\theta$:
\begin{equation}
    {\theta ^*} = \arg \mathop {\min }\limits_\theta  {\mathbb{E}_{(x,y) \in {{\cal D}_{{\rm{pseudo}}}}}}[{{\cal L}_{{\rm{sup}}}}({F_\theta }(x),y)],
\end{equation}
where $\theta ^*$ represents the final optimized model parameters.

After full parameter fine-tuning, the model effectively adapts to new data features and optimizes their performance in real-world scenarios.

\subsection{Hard Patch Selection Algorithm}
\begin{figure*}[!ht]
\centering
\includegraphics[width=1\textwidth]{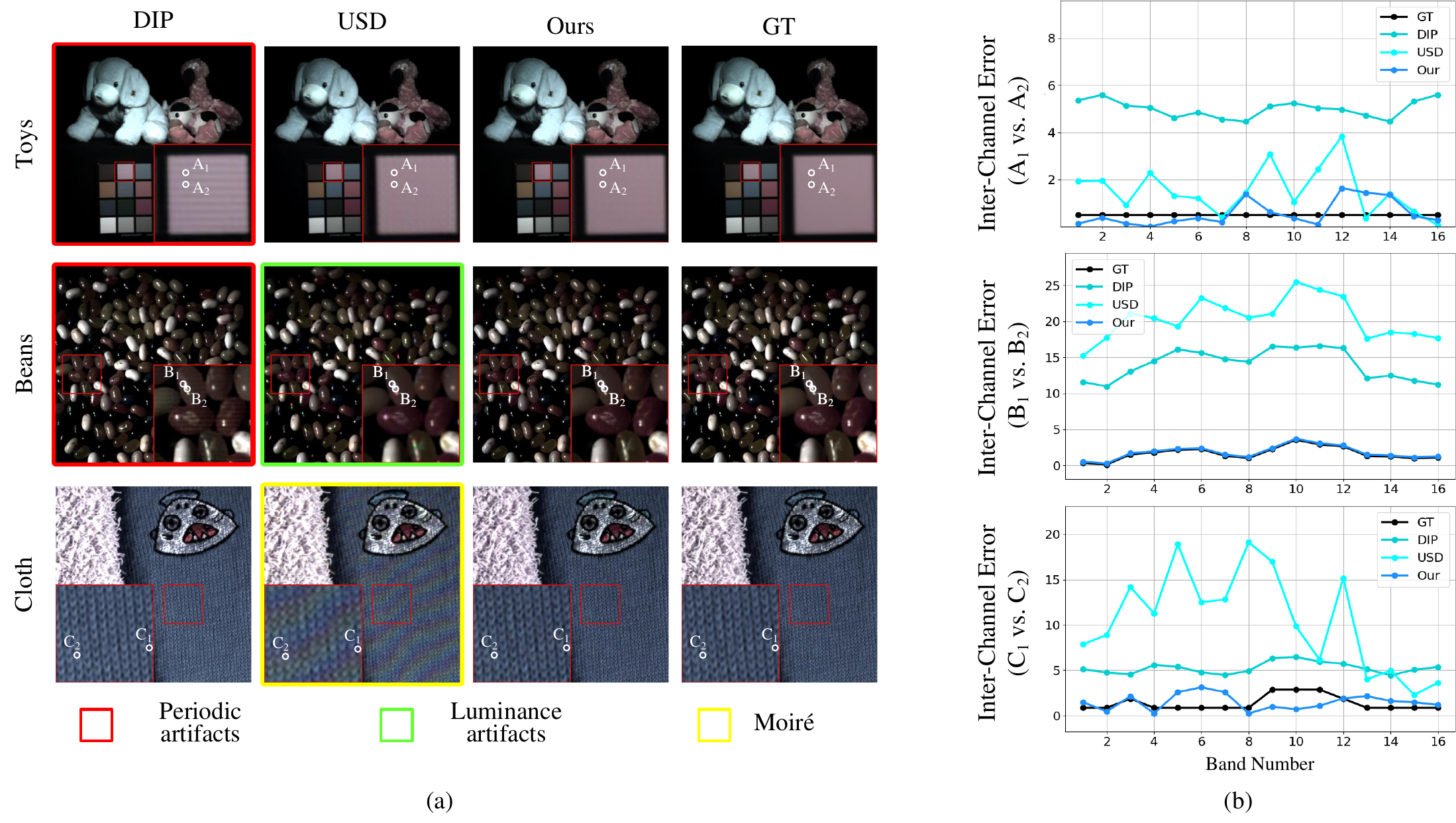} 
\caption{Comparison of GMAD (with MCAN+ as the backbone) with other SOTA demosaicing methods. (a) Three types of artifacts are typically produced during the demosaicing process, indicated by red, green, and yellow boxes, respectively. (b) Comparison of spectral curves at different points at the artifacts. It can be seen that the artifacts correspond to larger errors in the spectral curves, and the larger the jitter in the error curves, the larger the difference in the corresponding spectral features.} 
\label{Curve_figure}
\end{figure*}

During our experiments, we found that existing deep learning methods are prone to introducing artifacts in the process of demosaicing. These artifacts can be mainly categorized into three types: periodic artifacts, luminance artifacts, and moiré. The visualization of these three artifacts is shown in Fig. \ref{Curve_figure}(a), which reflect an incorrect reconstruction of the original spectral signal.

To further analyze the effect of artifacts on the accuracy of spectral signal reconstruction, we performed a comparative analysis of the spectral feature curves of different methods at multiple pixel points of the same object, and the results are shown in Fig. \ref{Curve_figure}(b). Under ideal conditions, the spectral characteristics of a homogeneous object should remain consistent across different pixel points \cite{liang2018material}, as these characteristics are intrinsic properties of the material. However, as can be seen from the figure, the spectral curves of the artifact generation method show obvious deviations between neighboring pixel points. They are manifested as inconsistent peak and trough positions, which in turn leads to the wrong characterization of the material properties. In contrast, the spectral characterization curves of other points are much closer to the GT, and the overall trend remains consistent, showing higher reliability. This suggests that artifacts not only affect the visual quality of the demosaicing results but also significantly reduce the accuracy of the spectral features, thus weakening the ability to recognize substances based on spectral information. Therefore, reducing the generation of artifacts is crucial for improving the accuracy of spectral analysis. 
Although traditional global mean error metrics (e.g., PSNR) are widely used in evaluating image reconstruction performance, such metrics are difficult to effectively reflect the presence and distribution of artifacts in an image \cite{nowScoreBad} because some of the artifacts only occupy a small pixel area. 

\begin{figure*}[ht]
\centering
\includegraphics[width=1\textwidth]{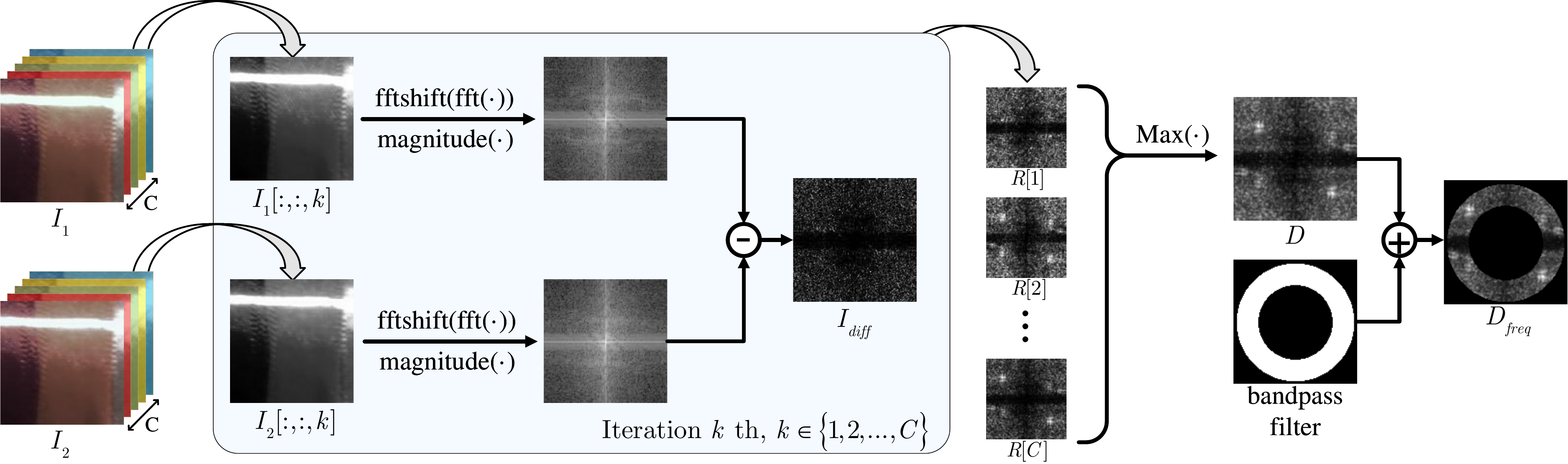} 
\caption{The process of generating a frequency variation map. It is traversed k times and processed step by step for each channel. Then the maximum value is taken for all pixel points, highlighting the frequency components with the largest variations. Finally, a bandpass filter is used to filter out non-artifactual parts of the frequency band.}
\label{Dfreq}
\end{figure*}

\textbf{Generate Frequency Variation Map.}
To detect and localize artifacts more accurately, we design a method of artifact detection and hard patch selection based on frequency domain analysis, the core idea of which is to observe the spectral disparity map through the Fourier transform to capture potential artifact features in the demosaicing results. To illustrate, given two input images ${I_1},{I_2} \in {^{H \times W \times C}}$ and bandpass filter parameters ${r_{low,}}{r_{high}}$,  the generation process is described as follows: 

(1) Channel-Wise Fourier Transform. For each spectral channel $k\in \{ 1,2,...,C\} $, compute the two-dimensional Fourier transform: 
\begin{equation}
    {D_1} = fft({I_1}[:,:,k]){\rm{, }}\ {D_2} = fft({I_2}[:,:,k]).
\end{equation}
Then, the DC component was then moved to the center of the spectrum:
\begin{equation}
    {S_1} = fftshift({D_1}){\rm{, }}{S_2} = fftshift({D_2}).
\end{equation} 

(2) Logarithmic Amplitude Difference. To facilitate a direct comparison of magnitudes, take the logarithm of the amplitude (where $\epsilon$ is a small constant to avoid numerical overflow):
\begin{equation}
    {M_1} = \log (\left| {{S_1} + \epsilon} \right|){\rm{, }}{M_2} = \log (\left| {{S_2} + \epsilon} \right|).
\end{equation}

Next, compute the difference in their magnitudes: 
\begin{equation}
    {I_{diff}} = \left| {{M_1} - {M_2}} \right|.
\end{equation} 
To mitigate noise, perform Gaussian smoothing on ${I_{diff}}$, yielding: 
\begin{equation}
    R[k] = GaussianBlur({I_{diff}}).
\end{equation}

(3) Maximum Across Channels and Bandpass Filtering. After processing all channels, take the pixel-wise maximum of $R[k]$ across channels to highlight the frequencies with the greatest differences: 
\begin{equation}
D = \mathop {\max }\limits_k \{R[k]\}.
\end{equation}
Finally, apply a bandpass filter to remove irrelevant low-frequency and very high-frequency components while preserving mid-to-high frequency content potentially indicative of artifacts. This yields the final frequency-domain difference map: 
\begin{equation}
    {D_{freq}} = {\mathop{\rm filter}\nolimits} (D,{r_{low,}}{r_{high}}).
\end{equation}

\textbf{Select Hard Patches.}
After generating the frequency variation map and obtaining the detection results, we further design a strategy based on threshold determination for constructing the hard patch dataset. Specifically, we first set the frequency intensity threshold $T_{{\mathop{\rm var}} }$, and mark the pixel points whose intensity exceeds this threshold in the frequency variation map by $mask$:
\begin{equation}
    mask(u,v) = \left\{ {\begin{array}{*{20}{l}}
        {1{\rm{, if\ }}{{\cal D}_{freq}}(u,v) > {T_{{\mathop{\rm var}} }}}\\
    {0{\rm{, otherwise}}}
\end{array}} \right..
\end{equation}
Calculate the total number of pixels above the threshold $count$: 
\begin{equation}
    count = \sum\limits_{x = 1}^H {\sum\limits_{y = 1}^W {mask(x,y)} }, 
\end{equation}
and compare it with the preset total number of pixels threshold ${T_{cnt}}$ to determine whether the patch is a hard patch: 
\begin{equation}
    isHard = \left\{ {\begin{array}{*{20}{l}}
    {True{\rm{, if\ count}} > {T_{cnt}}}\\
    {False{\rm{, otherwise}}}
    \end{array}} \right..
\end{equation}
With the above strategy, we filter out all the hard patches to construct the dataset ${\cal D}_{hard}$: 
\begin{equation}
    {{\cal D}_{hard}} = \{ {I_i}|isHard({I_i}) = True,\forall i \in {{\cal D}_{pseudo}}\}.
\end{equation}

\textbf{Fine-Tuning Based on Hard Patches.}
After completing the construction of the hard patches dataset, we perform the fine-tuning training in step 3 on ${\cal D}_{hard}$ to further improve the demosaicing performance of the model and finally obtain the optimized optimal parameters ${\theta^* }$: 
\begin{equation}
    {\theta ^*} = \arg \mathop {\min }\limits_\theta  {\mathbb{E}_{(x,y) \in {{\cal D}_{hard}}}}[{{\cal L}_{{\rm{sup}}}}({F_\theta }(x),y)].
\end{equation}
By selecting hard patches with significant artifacts in the frequency domain and performing targeted fine-tuning of the model, we effectively reduce the generation of artifacts during the demosaicing process and improve the spectral consistency and accuracy of the reconstruction results. This approach is inspired by the hard example mining technique in the field of target detection \cite{hardMining1, hardMining2} and the moiré image selecting method \cite{MIT, Fourier}.

\subsection{The Proposed Dataset: UniSpecTest}

\begin{figure}[ht]
\centering
\includegraphics[width=0.47\textwidth]{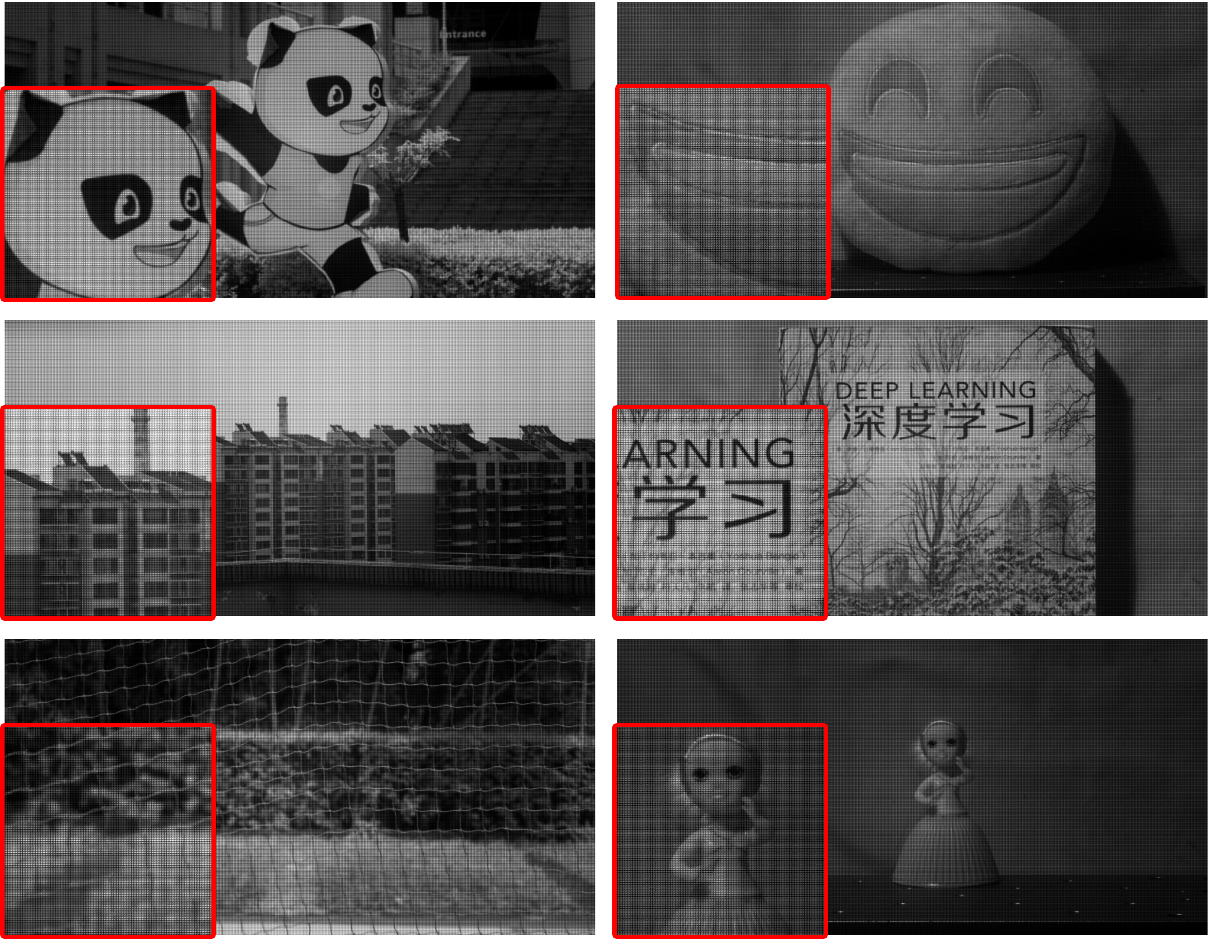} 
\caption{The proposed UniSpecTest, a multispectral mosaic test set containing diverse subjects, materials, and lighting conditions.}
\label{UniSpecTest}
\end{figure}

\begin{figure*}[ht]
\centering
\includegraphics[width=1\textwidth]{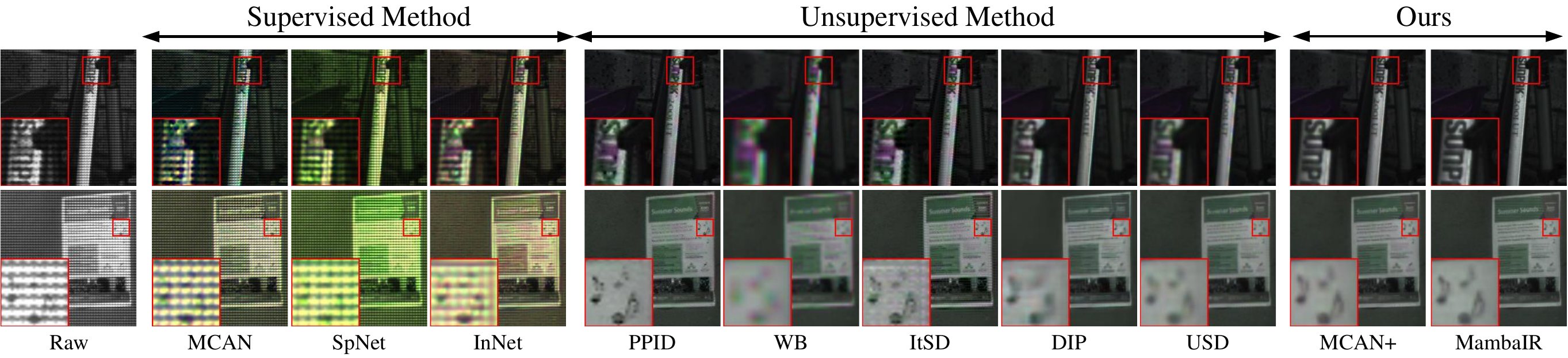}
\caption{Comparison of visual performance of various supervised and unsupervised methods on the Stereo dataset. Our method not only significantly reduces artifacts when handling complex textures and edges.}
\label{compare_Stereo}
\end{figure*}
\begin{figure*}[h]
\centering
\includegraphics[width=1\textwidth]{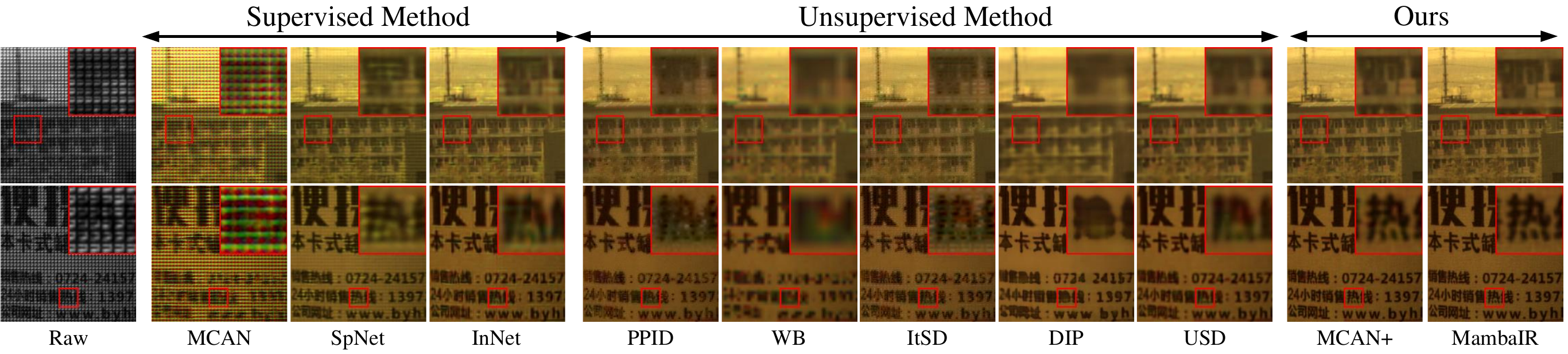}
\caption{Comparison of visual performance of various supervised and unsupervised methods on the UniSpecTest dataset. Compared to other methods, our approach demonstrates the best detail restoration in real-world scenarios.}
\label{compare_UniSpecTest}
\end{figure*}

To further validate the performance of our method under real-world conditions, we introduced the multispectral image test set UniSpecTest. The test set consists of 38 mosaic images, each with 25 spectral bands and a resolution of 2045$\times$1080, as demonstrated in Fig. \ref{UniSpecTest}. These images cover a wide range of objects, such as outdoor transportation, natural landscapes, buildings, road signs, and reflectors, as well as indoor pillows, plastic toys, and books. These objects are diverse in terms of materials and lighting conditions, and all of them were captured by an IMEC 25-band snapshot spectral camera in the spectral range of 600 nm to 900 nm. At the same time, this dataset can also provide a broad test benchmark for other studies.

\section{Experiments and Analysis}
\subsection{Datasets and Metric}

\begin{table*}[b!]
\caption{Dataset Details for Pre-Training and Fine-Tuning. For Datasets With Uniform Sizes, Spatial Resolution Is Expressed as “Height × Width”. The "Function" Row Indicates the Dataset Usage, with the Number of Images Specified in Parentheses.}
\centering
\begin{tabular}{ccccccc}
\toprule
\multirow{2}{*}{Datasets} &\multicolumn{2}{c}{Pretrain dataset}  & Real-world experiment  &\multicolumn{2}{c}{Real-world experiment} & Simulation experiment \\
\cmidrule(lr){2-7}
& ICVL & NTIRE & Stereo & Mosaic25 & UniSpecTest & CATT \\
\midrule
Function  & Training (40) & Training (900) & Training (168) & Training (50) & --- & Training (40) \\
          & Validation (10) & Validation (50) & Validation (48) & Validation (9) & --- & Validation (5) \\
          & Test (–)      & Test (–)      & Test (50)      & Test (–)      & Test (38) & Test (10) \\
Resolution & 1390$\times$1300 & 480$\times$512 & 256$\times$512 & --- & 2045$\times$1080 & --- \\
$N_{MSFA}$    & 5   & 4  & 4    & 5   & 5    & 4\\
\bottomrule
\end{tabular}
\label{datasets}
\end{table*}

\begin{figure*}[ht]
\centering
\includegraphics[width=1\textwidth]{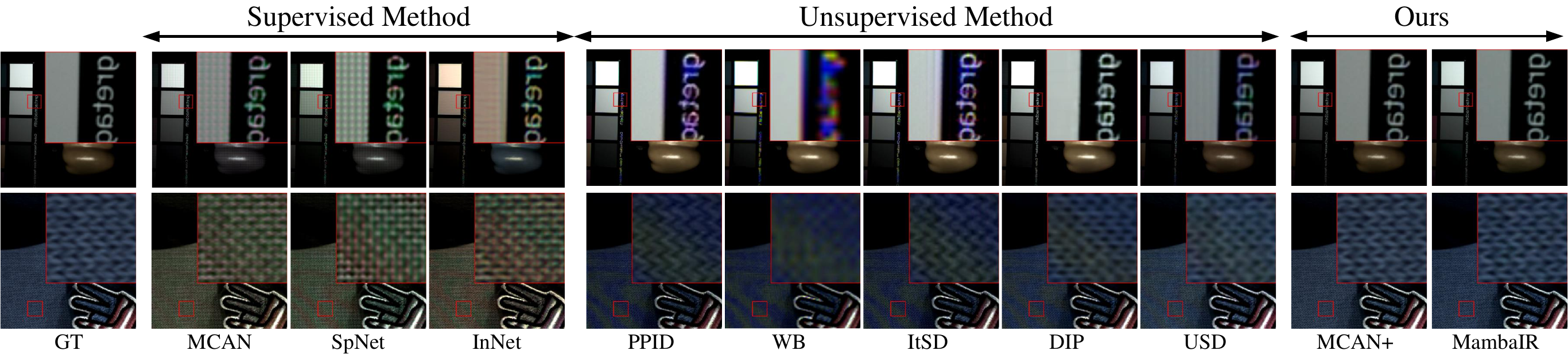}
\caption{Comparison of visual performance of various supervised and unsupervised methods on the CATT dataset. The results of our GMAD method are the closest to GT and produce the fewest artifacts while maintaining texture integrity.}
\label{compare_CATT}
\end{figure*}

We performed three fine-tuning experiments containing two sets of detail-rich multispectral data from real scenes and one set of synthetic datasets taken in the lab. We use average PSNR, SSIM, and SAM as the metrics. The detailed settings of the datasets are shown in Table \ref{datasets}.
\subsubsection{Real-World Dataset} In the first experiment, the Stereo data \cite{Stereo} contains common natural and office environments in life, as well as complex living objects with rich image details. In the second experiment, the training set for 5$\times$5 pattern is from Mosaic25 \cite{USD}, which was taken with the same camera as our proposed test set UniSpecTest.
\subsubsection{Simulation Dataset} To effectively measure the effect of demosaicing quantitatively, we used a mixture of the CAVE dataset \cite{CAVE} and the TT31 dataset \cite{TT31}, which is referred to as the CATT dataset. Both were made with the same type of camera, and both were shot under the CIE standard light source D65.
\subsubsection{Pretrain Dataset}In the pre-training stage, we used the simulated datasets NTIRE dataset \cite{NTIRE} and ICVL \cite{ICVL} dataset corresponding to the two mosaic patterns. 

\subsection{Implementation Details}
DIP is a classical generative approach that, unlike diffusion-based methods, does not require training on large-scale image datasets. Therefore, to facilitate the validation of our proposed method, we employed DIP to generate pseudo labels.

In our study, we use two different backbones for experimental comparison: the MCAN+ and the MambaIR, where the MCAN+ is based on the original MCAN model and facilitates inter-method comparison by increasing the number of MRAB modules to match the size of other CNN-based methods.

For the training set, we used a data enhancement method that flipped and rotated horizontally or vertically by 90$^{\circ}$, 180$^{\circ}$ and 270$^{\circ}$. During training, we cropped the 4$\times$4 pattern of data into blocks of size 128$\times$128 and the 5$\times$5 pattern of data into blocks of size 100$\times$100. The optimizer is Adam \cite{Adam}. Empirically, the learning rate was set $1\times10^{-5}$ for pre-training and $1\times10^{-4}$ for fine-tuning. The learning rate was multiplied by 0.5 per 1000 epochs. All the experiments were performed using the Pytorch framework on an NVIDIA RTX4090 GPU server. 3,000 rounds of pre-training and 5,000 rounds of fine-tuning were performed to select the model with the best validation loss as the best model for testing.

\subsection{Comparison With State-of-the-Art Methods.}
We use models fine-tuned on hard patches as comparison models. The six methods used for comparison include WB \cite{WB}, ItSD \cite{ItSD}, PPID \cite{PPID}, DIP \cite{DIP}, MCAN \cite{MCAN}, and USD \cite{USD}. To be fairly comparable at the time of the comparison, the supervised methods MCAN, SpNet, and InNet are pre-trained on the NTIRE dataset, and the unsupervised method USD is pre-trained on the corresponding unlabeled training set on each dataset, taking the models that converged during training as a comparison.

The visual demosaicing results of various methods across three datasets, including both simulated and real-world datasets with 4 $\times$ 4 and 5 $\times$ 5 mosaic patterns, are presented as follows.

\textbf{Performance on Stereo Dataset:}
As shown in Fig. \ref{compare_Stereo}, traditional supervised methods such as MCAN, SpNet, and InNet, which rely on features learned from simulated camera data, often struggle to effectively eliminate periodic artifacts. 
Although the unsupervised method USD shows better adaptability to new data, it still produces significant artifacts in real-world scenes. 
In contrast, our method not only significantly reduces artifacts when handling complex textures and edges but also generates the sharpest and clearest edge in character and note patterns.

\textbf{Performance on UniSpectest Dataset:}
Fig. \ref{compare_UniSpecTest} illustrates that architectural elements pose additional challenges for unsupervised methods, leading to blurred edges and obscure details.
The weak constraints of unsupervised methods often result in fine-grained structure loss, leading to suboptimal reconstructions.
In contrast, GMAD successfully preserves sharp edges, retaining intricate details and reducing unwanted artifacts.

\textbf{Performance on CATT Dataset:}
In Fig. \ref{compare_CATT}, we can observe that fabric textures present the most challenging case, where many methods generate unnatural line patterns that deviate significantly from the true texture.
The supervised MCAN, SpNet, and InNet approaches exhibit limitations in handling such fine-grained textures, resulting in noticeable distortions.
The superior performance of GMAD in these scenarios highlights its ability to effectively minimize artifacts while maintaining texture integrity.

Table \ref{compare_table} demonstrates that our method consistently outperforms all other methods across all three metrics, indicating that the proposed GMAD provides significant improvements in image quality, structural accuracy, and spectral recovery. A key advantage of our method is that it does not require GT, making it ideal for real-world snapshot spectral imaging applications where labeled data is not available.

\begin{table*}[!ht]
\caption{Quantitative Test Results on the CATT Dataset. The \textbf{bold} and \underline{underline} Indicate the Best and Second-Best Results, Respectively. The Proposed Demosaicing Method Is Far Superior to Existing Methods in All Three Metrics}
\centering
\begin{tabular}{rcccccccccc}
\toprule
\multirow{2}{*}{Metrics} &\multicolumn{3}{c}{Supervised} &\multicolumn{5}{c}{Unsupervised} &\multicolumn{2}{c}{Ours} \\
\cmidrule(lr){2-4} \cmidrule(lr){5-9} \cmidrule(lr){10-11}
&MCAN&SpNet&InNet&WB&ItSD&PPID&DIP&USD&MCAN+&MambaIR\\
\midrule
PSNR$\uparrow$&29.6364&28.7414&28.4088&30.4964&37.4648&37.5827&40.5903&43.4248&\underline{46.1486}&\textbf{46.5247} \\
SSIM$\uparrow$&0.8158&0.8246&0.8655&0.8866&0.9657&0.9791&0.9814&0.9904&\underline{0.9947}&\textbf{0.9972} \\
SAM$\downarrow$&10.0083&9.2083&9.572&5.3699&5.0657&3.6051&3.5362&3.2609&\underline{2.22}&\textbf{2.0315}\\
\bottomrule
\end{tabular}
\label{compare_table}
\end{table*}

\begin{table*}[h]
\caption{Quantitative Comparison of Three Patch Selection Strategies for Fine-Tuning on the CATT Dataset. The Three Indicators of Fine-tuning on Hard Patches Outperforms Others. This Indicates That Focusing on Difficult Data Improves Model Generalization}
\centering
\begin{tabular}{cccccccc}
\toprule
\multirow{2}{*}{Metrics} & \multicolumn{3}{c}{MCAN+} & \multicolumn{3}{c}{MambaIR} \\
\cmidrule(lr){2-4} \cmidrule(lr){5-7}
& Random patches& Hard patches& All patches& Random patches & Hard patches& All patches\\
\midrule
PSNR $\uparrow$ & 45.1059 & \textbf{46.1486} & 45.4716	&	45.5631 & \textbf{46.5247} & 45.7427   \\
SSIM $\uparrow$ & 0.9916 & \textbf{0.9947} & 0.9919  & 0.9951 & \textbf{0.9972} & 0.9952 \\
SAM $\downarrow$ & 2.6491 & \textbf{2.2200} & 2.5789  & 2.4963 & \textbf{2.0315} & 2.4360 \\
\bottomrule
\end{tabular}
\label{woHardpatch_table}
\end{table*}

\subsection{Ablation Study}
\subsubsection{The Effectiveness of Pseudo-Paired Datasets}
We present a comparison of fine-tuning results using the real paired dataset versus the pseudo-paired dataset generated by the DIP model, as shown in Table \ref{DIPorGT_table} and Fig. \ref{DIPorGT_teaser_figure}. The results highlight the close performance between fine-tuning on the pseudo-paired dataset and the real labels, with notable observations across both PSNR and SAM metrics.

In particular, when fine-tuning the MCAN+ network with pseudo-paired data generated using DIP, its SAM score surpasses that of the fine-tuning using GT. This suggests that the pseudo-paired data, although generated in an unsupervised manner, provides results comparable to real data. 

The superior performance of pseudo-paired data can be attributed to several factors. The feature richness captured through pseudo-labeling likely enhances the model’s generalization capability. This indirect supervision strategy encourages the network to learn more broadly applicable features for demosaicing, allowing it to perform well across a range of previously unseen data. The improved generalization helps the network avoid overfitting to the limited real paired data and allows it to develop a more robust demosaicing model.

\begin{table}[h]
\caption{Comparison of Fine-Tuning with Pseudo-Paired Data and GT on the CATT Dataset. On MCAN+ and MambaIR Networks, the Results of Using Pseudo Label Fine-Tuning are Very Close to GT Fine-Tuning Results and Even Exceed in a SAM Index, Which Fully Verifies the Effectiveness and High Performance of GMAD}
\centering
\begin{tabular}{cccccc}
\toprule
\multirow{2}{*}{Metrics} & \multicolumn{2}{c}{MCAN+} & \multicolumn{2}{c}{MambaIR} \\
\cmidrule(lr){2-3} \cmidrule(lr){4-5}
& Pseudo labels & GT & Pseudo labels & GT \\
\midrule
PSNR $\uparrow$ & 45.4716 & \textbf{45.54612} & 45.7999 & \textbf{47.1725}  \\
SSIM $\uparrow$ & 0.9929 &	\textbf{0.9939} & 0.9972 & \textbf{0.9974}\\
SAM $\downarrow$ & \textbf{2.5789} & 2.6185 & 2.0315 & \textbf{1.9959}\\
\bottomrule
\end{tabular}
\label{DIPorGT_table}
\end{table}

\subsubsection{The Significance of Pre-training}

\begin{table}[h]
\caption{Comparison of With(W/) or Without(W/O) Loading Pre-Trained Models. The Data Show That Loading a Pre-Trained Model Improves Model Generalization and Thus Performance on New Datasets
}
\centering
\begin{tabular}{cccccc}
\toprule
\multirow{2}{*}{Metrics} & \multicolumn{2}{c}{MCAN+} & \multicolumn{2}{c}{MambaIR} \\
\cmidrule(lr){2-3} \cmidrule(lr){4-5}
& W/O & W/ & W/O  & W/ \\
\midrule
PSNR $\uparrow$ & 42.4501 & \textbf{45.4716}	& 44.4929 &	\textbf{45.7999}  \\
SSIM $\uparrow$ & 0.9887 &	\textbf{0.9929} & 0.9929 & \textbf{0.9972}\\
SAM $\downarrow$ & 3.5528 & \textbf{2.5789} & 2.5357 & \textbf{2.0315}\\
\bottomrule
\end{tabular}
\label{woPretrain_table}
\end{table}

To assess the impact of pre-training in the multispectral domain, we conducted a comparative experiment where one method was trained directly on the CATT dataset, while the other was first pre-trained on a large scale simulated dataset before fine-tuning on the CATT dataset. The quantitative results, as shown in Table \ref{woPretrain_table}, indicate that all supervised methods, when fine-tuned with a pre-trained model, outperform those trained directly on the target dataset. This demonstrates the advantages of pre-training in enhancing model generalization and performance on real-world scenes.

Notably, the MCAN+ method exhibits the most significant improvement, with a 3.02 dB increase in PSNR and a 0.9739 decrease in SAM. These improvements highlight that pre-training allows the model to better recover the fine-grained spectral details and original features of the multispectral data, leading to higher quality demosaicing outputs. The PSNR and SAM metrics show that pre-training not only improves the signal-to-noise ratio but also enhances the structural accuracy of the demosaiced images, reducing undesirable artifacts and preserving critical details.

Furthermore, MambaIR also shows impressive results, particularly in terms of SSIM. This indicates that MambaIR has a significant advantage in preserving the brightness, contrast, and structural similarity in demosaicing, making it a highly effective model for real-world applications where these factors are crucial.

\begin{figure*}[h!]
\centering
\includegraphics[width=1\textwidth]{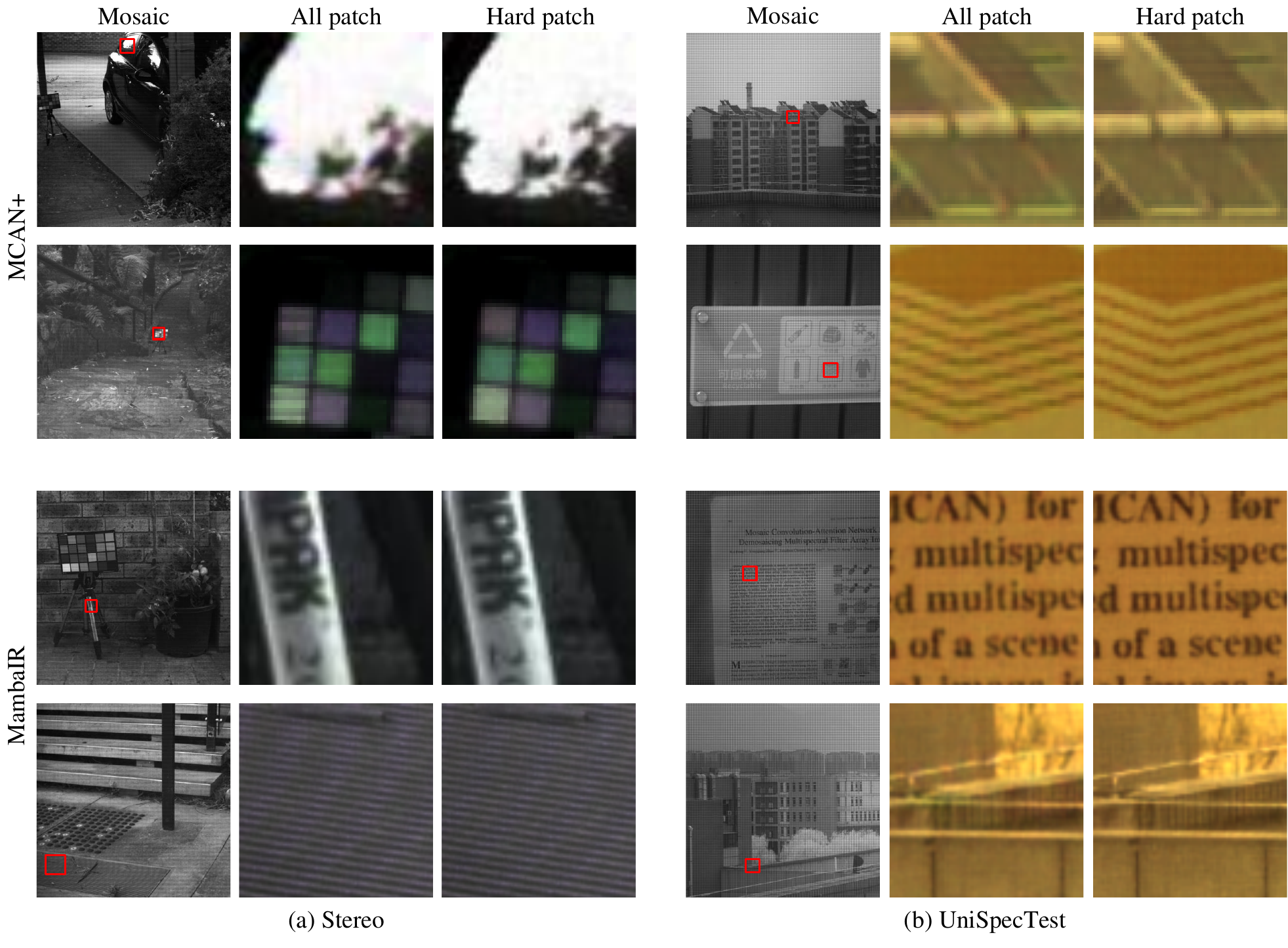} 
\caption{Quantitative Comparison of Three Patch Selection Strategies for Fine-Tuning on the CATT Dataset. On Two Real-World Datasets, the Artifacts on Hard Patches Are More Slight Compared to Those on All Patches.}
\label{woHardpatch_figure}
\end{figure*}

\subsubsection{The Significance of Selecting Hard Patches}
To evaluate the effect of fine-tuning with hard patches, we employed three patch selection strategies on the CATT dataset. The first strategy used all patches for fine-tuning, the second strategy fine-tuned using only the selected hard patches, and the third strategy involved randomly selecting patches equal in number to the hard patches for fine-tuning. The visual comparison of these strategies is shown in Fig. \ref{woHardpatch_figure}. As can be seen from the figure, the images fine-tuned with hard patches exhibit fewer periodic artifacts, luminance artifacts, and moiré patterns.

In addition, we performed a quantitative analysis of the fine-tuning performance, as shown in Table \ref{woHardpatch_table}. The three metrics clearly demonstrate that fine-tuning with hard patches yields the best performance, followed by the use of all patches, while the random patches strategy results in the least favorable outcome. This ranking suggests that focusing on hard patches significantly enhances the model's ability to recover fine details and suppress artifacts, ultimately improving overall demosaicking performance. By focusing on the challenging regions that present difficulties for the network, the model is better able to handle complex patterns in real-world data, thereby improving its robustness and spectral fidelity.
 
\section{Conclusion}
In this paper, we propose GMAD, a novel training framework for hybrid supervised multispectral demosaicing in scenarios where real-world datasets lack GT. GMAD adopts a three stage training strategy: (1) \textbf{pre-training} on simulated data to learn robust spectral features; (2) \textbf{pseudo-pairing} using the Deep Image Prior generative model to produce pseudo demosaiced cubes; and (3) \textbf{fine-tuning} the network guided by the pseudo demosaiced cubes generated in Stage 2. To enhance fine-tuning efficiency, we introduce a hard patch evaluation and filtering mechanism, which significantly reduces artifacts by focusing on high quality training samples. Additionally, we contribute UniSpeTest, a new real-world mosaic test set for benchmarking. Extensive experimental results demonstrate that GMAD offers a robust and high performance solution for real-world multispectral demosaicing tasks.


\bibliographystyle{IEEEtran}
\bibliography{mylib}

\vfill

\end{document}